# AN IMPROVED DECENTRALIZED APPROACH FOR TRACKING MULTIPLE MOBILE TARGETS THROUGH ZIGBEE WSNS

Tareq Alhmiedat[1], Amer O. Abu Salem[2], Anas Abu Taleb[3]

[1]Department of Computer Science, Zarqa University, Zarqa, Jordan
`t.alhmiedat@zu.edu.jo`
[2]Department of Computer Information Systems, Zarqa University, Zarqa, Jordan
`abusalem@zu.edu.jo`
[3]Department of Computer Science, Isra University, Amman, Jordan
`anas.taleb@ipu.edu.jo`

## ABSTRACT

*Target localization and tracking problems in WSNs have received considerable attention recently, driven by the requirement to achieve high localization accuracy, with the minimum cost possible. In WSN based tracking applications, it is critical to know the current location of any sensor node with the minimum energy consumed. This paper focuses on the energy consumption issue in terms of communication between nodes whenever the localization information is transmitted to a sink node. Tracking through WSNs can be categorized into centralized and decentralized systems. Decentralized systems offer low power consumption when deployed to track a small number of mobile targets compared to the centralized tracking systems. However, in several applications, it is essential to position a large number of mobile targets. In such applications, decentralized systems offer high power consumption, since the location of each mobile target is required to be transmitted to a sink node, and this increases the power consumption for the whole WSN. In this paper, we propose a power efficient decentralized approach for tracking a large number of mobile targets while offering reasonable localization accuracy through ZigBee network.*

## 1. INTRODUCTION

In recent years, advances in signal processing have led to manufacturing small, low power, inexpensive Wireless Sensor Network (WSN). The signal processing in WSN is different from the traditional wireless networks in two critical aspects: first, the signal processing in WSN is performed in a fully distributed manner, unlike in traditional wireless networks. Second, due to the limited computation capabilities of sensor networks, it is essential to develop an energy and bandwidth efficient signal processing algorithms.

A sensor network is composed of sensor nodes which are small in size, low in cost, and have short communication range. A sensor node usually consists of four sub-systems as follows:

1. Computing subsystem, which is responsible for functions such as execution of the communication protocols and control of sensors

2. A sensing subsystem, that is responsible for sensing the environmental characteristics, such as temperature and humidity

3. A communication subsystem, this consists of a short radio range used to communicate with neighbouring nodes

4. A power supply subsystem, which includes a battery source that provides energy to sensor node.

WSN technology is exciting with unlimited potentials in various application areas including: environmental, medical, military, transportation, homeland defence, crisis management, entertainment, and smart spaces [1, 2, 3, 4, 5]. Researchers have focused on diverse aspects of WSN, such as hardware design, routing, data aggregation and localization [6, 7]. Recently, WSN-based localization and tracking issues have received much attention, driven by the need to accomplish high localization accuracy with the minimum cost, this is because:

1. In many applications, the location itself is the information of interest,

2. Several routing protocols are based on the sensor nodes' locations,

3. Transferring sensors' measurements without incurring the sensors' locations is an unproductive task.

The authors of this paper focused on diverse aspects of tracking mobile targets through distributed sensor networks, such as the localization method [8], tracking multiple mobile targets through ZigBee networks [9], and the data aggregation and prediction method [10, 11]. In this paper, we focus on the communication between nodes when tracking a large number of mobile targets through ZigBee WSNs, which further intends to reduce the power consumption for WSNs.

WSNs based Localization systems have been researched and addressed extensively in several works [12, 13, 14]. In real time tracking applications, it is essential to continuously transmit the mobile targets' locations to a sink node, in order to display its current position online. Tracking a large group of mobile targets through hundreds of sensor nodes requires the transmission of the location for each mobile target from reference nodes to the sink node, which normally results in a series of hops through the network. Each of these hops increases the consumption of the limited energy, and therefore participates failures within the network as the energy of the reference nodes becomes increasingly impaired.

In this paper, we categorize the tracking systems into centralized and decentralized. In centralized tracking systems, localization information might be transmitted to a sink node to obtain the localization information, while in decentralized systems; localization information is obtained from each mobile target itself, and then transmitted to a sink node. Figure 1 depicts the main idea for both systems (centralized and decentralized). As presented, assume a mobile target is in the range of 4 reference nodes. In centralized systems, a total number of 4 reference messages would be transmitted to the sink node, whereas in the decentralized systems, only a single message that includes the current location of the mobile target, is transmitted to the sink node, as processing the mobile target's location would take place at the reference or the mobile target node itself.

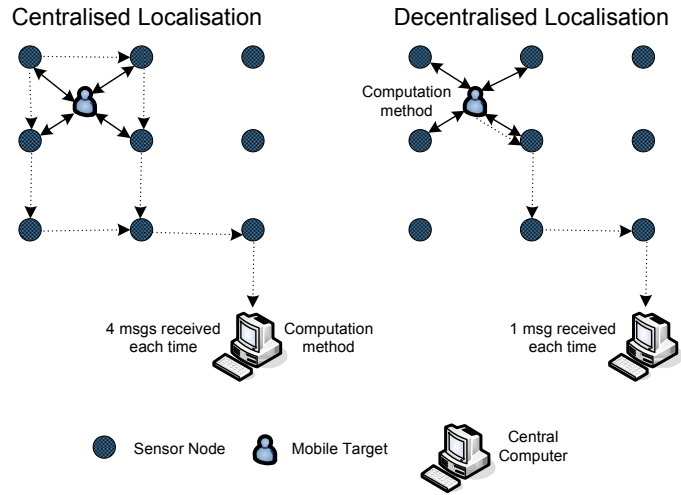

Figure 1. Centralized vs. decentralized localization methods

Tracking a few number of mobile targets through a decentralized approach offers low power consumption [10]. However, when it comes to tracking a large number of mobile targets through distributed sensor nodes, then high power consumption is expected. For that reason, this paper presents a new efficient tracking system, which basically reduces the power consumption required to track multiple mobile targets through ZigBee WSNs. In this paper, we will refer to the following terms:

- *Reference node (r)*: the node with fixed known 2-d coordinates,

- *Mobile target (t)*: the node with unknown location, which requires estimating the location of itself,

- *Group leader (g)*: the *mobile target node* with the minimum number of hops to a sink node is selected to be a group leader in order to aggregate the localization information obtained from adjacent mobile targets, and transmit to a sink node.

In this paper, our contribution lies on the following aspects:

1. Study the existing tracking systems, and categorize them into: centralized and decentralized systems,

2. Design, develop, and implement a power efficient decentralized tracking system using ZigBee WSNs, which intends to minimize the power consumption when tracking a large number of mobile targets,

3. Evaluate the proposed system in this paper, and compare it with the existing systems proposed recently.

## 2. RELATED WORK

In this section, we summarize the existing centralized and decentralized tracking systems. First, we start with centralized tracking systems. According to [15], centralized based systems have two advantages over the decentralized ones: first, reference nodes are not required to have processing capabilities. Second, the irregularity and mobility of the mobile target requires frequent data communications among reference nodes to have good localization accuracy, which destroys the main advantage of decentralized localization systems.

The works presented in [16, 17, 18] include a RSS based localization system through centralized communication among sensor nodes. In such systems, localization information is transmitted to a sink node in order to process and compute the localization coordinates for each mobile target at the sink node. The work presented in [19] investigated the performance of centralized cooperative positioning algorithm, in which all information has to be collected at a central entity for positioning target nodes. The system proposed in [20] includes a novel computing architecture for WSNs to overcome the Kalman Filter drawbacks, which divides the positioning system into three components: measurement, pre-processing, and data processing.

Second, decentralized or distributed localization techniques; which include that each sensor node is responsible for determining its location with only limited communication with nearby nodes. Decentralized systems have been widely used due to the low energy required compared to the centralized systems. However, this kind of systems requires attaching a localization and computation mechanism to each reference node. In [21], a hybrid technique called RObust Position Estimation (ROPE) was proposed, which allows sensors to locate their locations with no need to a centralized computation facility. ROPE provides a location verification mechanism that verifies the location claims of the sensors before data collection. Thus, ROPE allows the sensors to estimate their own location without the assistance of a central authority.

The system proposed in [22] involves a MDS-MAP technique for calculating the positions of nodes with only basic information that is likely to be already available. MDS-MAP technique involves starting with the given network connectivity information, and an all-pairs shortest-paths algorithm is used to assess the distance between each possible pair of nodes. A decentralized tracking system based on a camera sensor device is proposed in [23] which intended to reduce the bandwidth requirements. The images captured are required to be processed at each camera sensor node with the objective of the extracting location of mobile targets. The proposed approaches in [24, 25, 26] are decentralized based localization systems, since all the communication and processing are undertaken in the sensor node itself.

## 3. AN IMPROVED DECENTRALIZED TRACKING APPROACH

In WSN systems, long lifetime requirement of sensor nodes has led us to find out new horizons for reducing the power consumption upon nodes. Consuming less energy in tracking applications is a primary objective in designing WSN systems, as each sensor node is usually supported by batteries which could be difficult to replace. The proposed decentralized approach includes grouping the mobile targets in a given area, and selecting a group leader for each group of mobile targets. The group leader is responsible for:

1. Collecting localization information from other mobile targets in its range,

2. Aggregating and transmitting the received localization information to a sink node.

The proposed system in this paper is based on decentralized localization computation. Thus, localization information is processed at the mobile target device, in which each mobile target has the capability to compute its current location using the system proposed in [6]. Only the final coordinates are transmitted to a sink node. This reduces the amount of messages transmitted over the network, and hence shrinks the power-consumption for each reference node. In this section, we discuss the proposed decentralized tracking system, which consists of 3 main phases; initialization, localization, and grouping. Figure 2 depicts the idea of the improved decentralized approach.

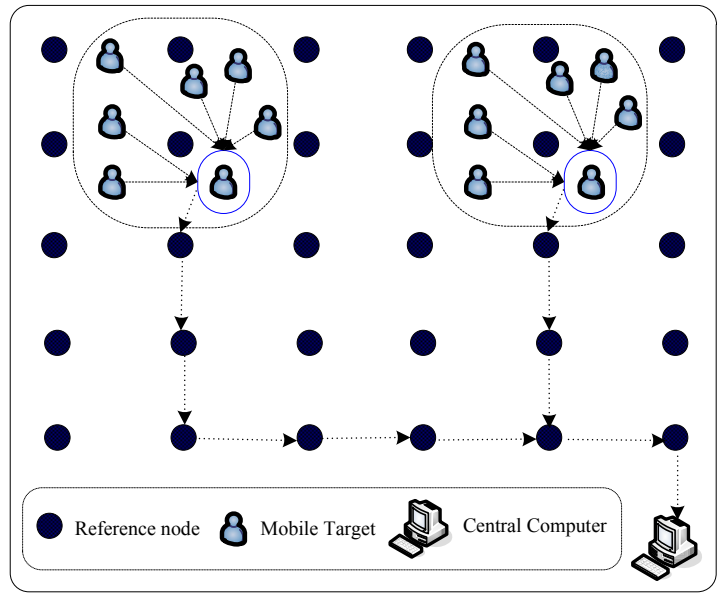

Figure 2. The main idea of the improved decentralized approach

### 3.1. Initialization phase

This phase includes starting the network, and assigning a network address to reference nodes in the tracking area of interest. Then, mobile targets enter the tracking area and obtain network addresses. Afterwards, each mobile target checks the reference nodes in its range, in order to position itself when required.

### 3.2. Localization phase

In this phase, each mobile target localizes itself based on the RSS values obtained from adjacent reference nodes. The localization approach proposed in [6] is deployed as a localization method to estimate the mobile targets' locations. This process is repeated every $f$ seconds to estimate the mobile targets' locations simultaneously.

### 3.3. Grouping phase

This phase involves selecting a group leader for the mobile targets set in the same transmission range. The main reasons behind choosing a group leader is to reduce the amount of messages transmitted to the sink node over the WSN, and hence conserving the amount of energy consumed during the tracking process.

To illustrate this phase, assume $R = \{r_1, r_2, r_3, ..., r_c\}$ represents the reference nodes in the tracking area of interest, with total number $c$, and $T = \{t_1, t_2, t_3, ..., t_e\}$ represents the mobile targets with total number $e$. Assume that $S = \{t_1, t_2, ..., t_z\}$ is a set of mobile targets in a given area set within the same transmission range, with a total number $z$, and each mobile target $t$ is covered by at least three reference nodes to triangulate its current position. Each mobile target checks the number of hops between itself and the sink node. The mobile target with the least number of hops to the sink node is elected to be a group leader. The group leader should have the following characteristics:

1. It has the minimum number of hops to the sink node,

2. It has a reasonable amount of energy, over a certain threshold.

The group leader receives the localization information from other mobile targets in its range, then aggregates, and transmits this information to the sink node. The flowchart for the algorithms running on a mobile target, and the group leader nodes are presented in Figures 3 and 4, respectively.

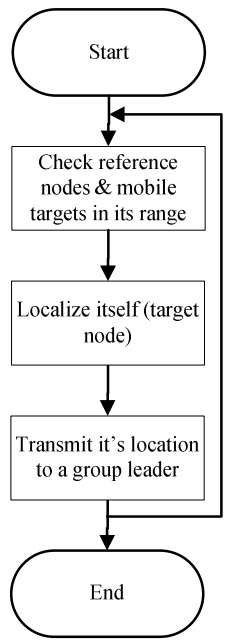

Figure 3. The algorithm implemented on each mobile target node

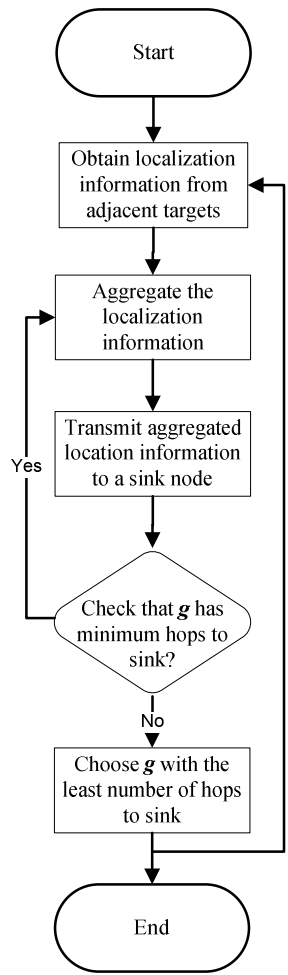

Figure 4. The algorithm implemented on the group leader node

## 4. SIMULATION EXPERIMENTS

The proposed system is evaluated through implementation using NS2 simulator environment. In order to evaluate the proposed approach in this paper, a total number of three different approaches have been implemented and tested. First, a centralized localization approach, which includes transmitting the localization information from each reference node covering the mobile target to a sink node. Second, a decentralized tracking system, this includes that each mobile target estimates its current position and transmits it to a sink node. And third, an improved decentralized approach, which involves selecting a group leader which placed close to the sink node in order to transmit the localization information through the minimum number of hops.

The system proposed in this work was implemented using ZigBee network standard. ZigBee is a low power consumption, low data rate, and low cost wireless communication standard designed to be used in home automation and remote control applications. ZigBee network standard performs three roles: coordinator, router, and end-device. Only one ZigBee coordinator is required for each WSN, it starts a network formation. ZigBee router is an optional network component; it participates in multi-hop routing of messages. While ZigBee end-device is utilized for low power operation, and is not allowed to participate neither in association nor routing. In this project, we used a ZigBee coordinator as a sink node and a number of ZigBee routers as reference and mobile target nodes.

## 4.1. Reference and mobile target platform

There are two different types of devices in our test-bed: reference nodes and mobile targets. Reference nodes include stationary sensor nodes with known positions distributed over the tracking area of interest, whereas the mobile targets have the ability to localize themselves. Both, reference and mobile targets were considered as router nodes. Table 1 presents simulation parameters used in our simulation experimental test-bed.

Table 1. Simulation parameters

| Parameter name | Parameter value |
|---|---|
| Packet size | 127 bytes |
| InitEng | 27.00 $mA.h$ |
| rxPower | 49.00 $mA.h$ |
| txPower | 44.00 $mA.h$ |
| Simulation time | 360 seconds |
| Number of nodes | 56 nodes |
| Number of mobile targets | 10 |
| Average hops | 5 |
| Radio model | TwoRayGround |
| Antenna type | OmniAntenna |
| Grid size | 75 x 65 $m^2$ |
| Routing protocol | AODV |
| Mac protocol | MAC/802.15.4 |

## 4.2. Reference and mobile target platform

Our experiment test-bed consists of 56 reference nodes distributed over the tracking area of interest, and a sink node placed on the right corner as depicted in Figure 5.

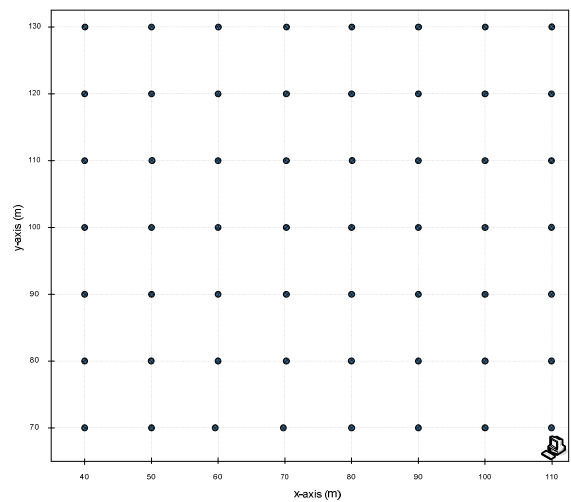

Figure 5. The experimental test-bed

## 5. SYSTEM EVALUATION

In this section, three scenarios are examined (centralized, decentralized, and improved decentralized), based on evaluating two performance metrics: the total number of messages transmitted over the network, and the power-consumption.

### 5.1. Communication cost

One of the critical issues in developing a tracking system for sensor nodes is the issue of battery lifetime. Sensor devices are designed to work for periods of several months to a few years. The system proposed in this paper aims to minimize the total number of messages transmitted over the network in order to reduce the amount of energy consumed. In this section, we evaluate the total number of messages transmitted to the sink node for the three different scenarios discussed above (centralized, decentralized, and improved decentralized).

In centralized localization approach, every reference node senses the mobile target has to transmit its readings to a sink node. The total number of messages transmitted over the network is represented in Equation 1.

$$n_1 = r \times m \times \frac{l}{f} \tag{1}$$

where $r$, $m$, $l$ and $f$ refer to the number of reference nodes, the number of mobile targets located in one transmission range, total experiment time, and frequency rate (how often the mobiles' locations have to reach the sink node) respectively. For instance, assume the total experiment time is $l = 60$ seconds, the number of reference nodes $r = 3$, and the number of mobile targets is $m = 10$, then the total number of messages transmitted to the sink node is $n_1 = 900$ (where $f = 2$ seconds).

On the other hand, in the basic decentralized approach, two types of messages are transmitted over the network. First, the messages transmitted between reference nodes and mobile targets represented in Equation 1. Second, the messages transmitted between mobile targets and a sink node, as represented in Equation 2.

$$n_2 = m \times \frac{l}{f} \tag{2}$$

As in the previous example, consider the total experiment time is $l = 60$ seconds, the number of reference nodes $r = 3$, and the total number of mobile targets is $m = 10$. Then the average number of local messages will be $n_2 = 900$. These messages are not required to reach a sink node, since they are required to be exchanged between reference nodes and mobile targets. Whereas the total number of messages sent to the sink node is 300 messages. Consequently, there is a significant improvement in the decentralized over centralized approaches.

The improved decentralized approach proposed in this paper, deals with three types of messages transmitted over the network, are as follows:

1. Local messages: indicates messages exchanged between each mobile target and reference nodes, which represented in Equation 1.

2. Group messages: represents messages transmitted between the group leader and other mobile targets in its range. This kind of messages does not affect the energy of reference nodes (wireless sensor nodes), as communications are processed between the mobile targets themselves, which represented in Equation 3.

3. Global messages: represents total number of messages transmitted between the group leader and the sink node, as represented in Equation 4.

$$n_3 = (m-1) \times \frac{l}{f} \tag{3}$$

$$n_4 = \frac{m}{5} \times \frac{l}{f} \tag{4}$$

In equation 4, the total number of messages is divided by 5, as each packet could contain the locations and network addresses for a maximum number of 5 mobile targets. For the same example mentioned above, assume the total experiment time is $l = 60$ seconds, the number of reference nodes $r = 3$, and the number of mobile targets is $m = 10$. Then the total number of local messages is $n_1 = 600$, the total number of group messages is $n_3 = 180$, and the total number of global messages is $n_4 = 40$.

As discussed above, the improved decentralized approach does not require a large number of messages to be exchanged between mobile targets and reference nodes, even when a large number of mobile targets are required to be positioned. Calculations are taken place at the mobile target side, with no need for centralized computations. Figure 6 shows a comparison of the messages required to be transmitted among the three scenarios. The centralized approach requires transmitting a high number of messages between reference and sink nodes, whereas the implemented decentralized approach requires exchanging less messages when the total number of mobile targets is small. At the meanwhile, the improved decentralized approach achieves the least number of messages required to be transmitted over the network mainly when the density of mobile targets is high.

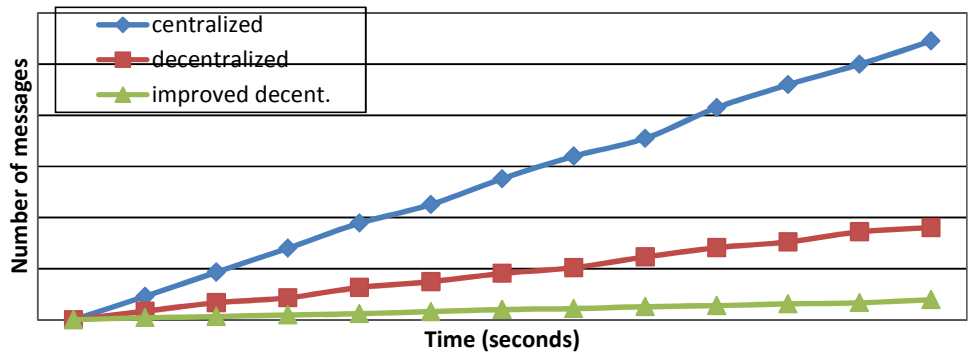

Figure 6. Total number of messages for the three systems

In this work, we aimed to reduce the amount of messages exchanged between wireless sensor nodes in the tracking area of interest. Figure 7 presents the total number of messages exchanged between mobile targets in the three scenarios. As presented, the improved decentralized approach achieves a few more number of messages than centralized and decentralized systems.

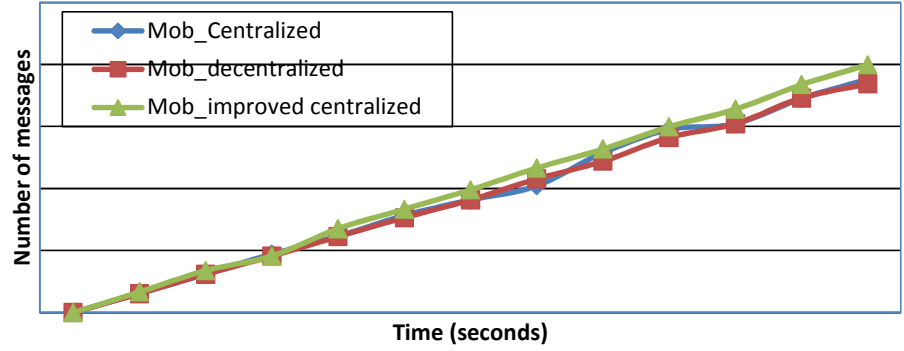

Figure 7. Total number of messages exchanged between mobile targets for the three systems

## 5.2. Power-consumption

The previous section evaluates and compares the total number of transmitted messages for the three scenarios. Each transmitted message requires a particular amount of power. In this section, the power-consumption for each approach is evaluated. The power consumption for the centralized $E_{cn}$, decentralized $E_{dc}$ and improved decentralized $E_{imp}$ represented in the Equations 8, 9, and 10, respectively.

$$P(n) = (n \times P_{tx}) + (n \times P_{rx}) \qquad (7)$$

$$E_{cn} = P(n_1) \times h \qquad (8)$$

$$E_{dc} = \left(P(n_1) + P(n_2)\right) \times h \qquad (9)$$

$$E_{imp} = \left(P(n_1) + P(n_4)\right) \times h \qquad (10)$$

where $P(n)$ is the energy required to transmit a number of packets $n$, $P_{tx}$ is the amount of energy required to transmit a single packet, $P_{rx}$ is the amount of energy required to receive a single packet, and $h$ is the number of hops. Figure 8 compares the battery life of the three approaches (centralized, decentralized, and improved decentralized). As shown, the improved decentralized method achieves longer battery life than both the centralized and decentralized approaches, as it requires less number of messages to be exchanged between mobile targets and a sink node.

In Figure 9, the power consumption for the group of mobile targets ($m = 10$) are estimated in three systems (centralized, decentralized and improved decentralized). As shown, the improved decentralized system achieves the highest power consumption. However, there is a slight difference in power consumption between the improved decentralized, centralized and decentralized systems. The mobile target devices are easy to maintain in case if any target device's battery has run out.

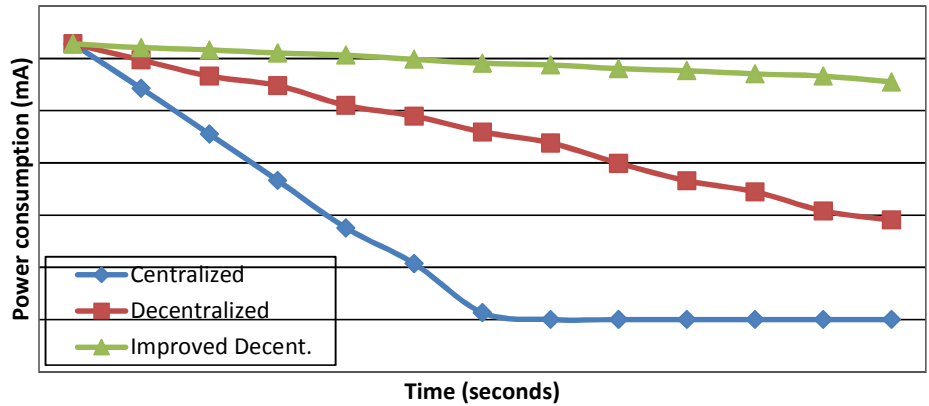

Figure 8. Evaluating the battery life for the reference nodes in the three systems

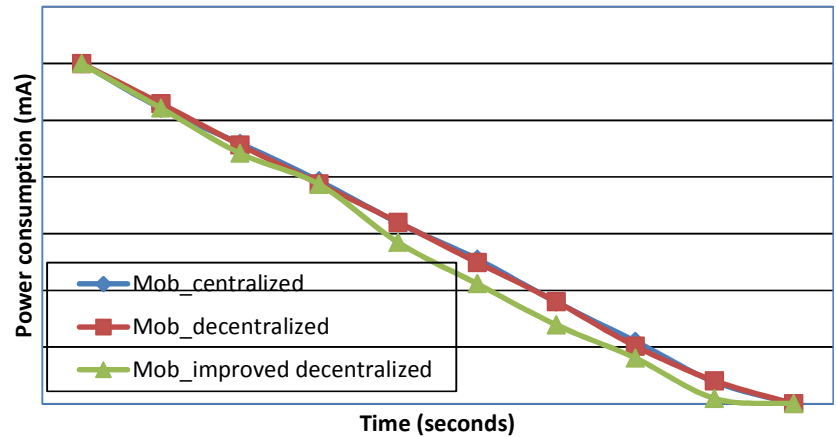

Figure 9. Evaluating the battery life for the mobile target nodes in the three systems

## 6. DISCUSSION

In WSN, energy has three main consumers: signal processing, data transmission, and hardware operations. In this research work, we aim to reduce the power consumption for both signal processing and data transmission. In the improved decentralized approach, reference nodes are not required to process the mobile targets' locations. Moreover, the total number of messages transmitted over the network is reduced.

Centralized based localization approaches [16, 17, 18] are much more power consuming than decentralized approaches [21, 22, 24, 25], since in centralized systems, usually the following details are required to be transmitted to a sink node: sensor id, target id, packet number and sensor-to-target distance. Therefore, deploying centralized systems with a large density of sensor nodes requires high power-consumption, as each localization message must be transmitted to a sink node via multi-hop network.

Decentralized tracking solutions are more attractive for large sensor networks including thousands of nodes, as there is no need to process the localization information at a sink node side, since all the computations are processed at each mobile target, and therefore the final coordinates are only requested to be transmitted to a sink node. The proposed decentralized systems in [21, 22, 24, 25] are efficient for tracking a single or a few number of mobile targets. Though, these systems are impractical in terms of power consumption to track a large number of mobile targets, as this requires a huge number of messages to be exchanged between mobile targets and the sink node, and this will increase both the power consumption and messages dropping rate. For both systems (centralized and decentralized), increasing the number of mobile targets will increase the number of transmitted messages to the sink node.

An improved decentralized approach is proposed in this paper in order to shrink the total number of messages transmitted over the network and hence decrease the power-consumption. In this system, reference nodes are not required to have high processing capabilities, since computations are taking place at each mobile target node. Grouping mobile targets which set/placed in the same transmission range will increase the tracking efficiency in terms of power consumption.

The proposed improved decentralized tracking system achieves low power consumption, and is applicable to track several mobile targets simultaneously with the minimum communication cost, in such applications necessitate tracking a large number of mobile targets. The system proposed in this paper can be deployed in different types of applications including: tracking a large number of employees in a workplace, tracking passengers in an airport, and tracking fire-fighters in hazard situations.

# 7. PAPER SUMMARY

One of the critical technical issues which must be addressed in developing sensor networks for object tracking applications is energy conservation. In this paper, a decentralized tracking approach is proposed to track the location of multiple mobile targets with minimum communication cost among sensor nodes. The proposed approach achieves an efficient tracking system in terms of localization and power-consumption by grouping the mobile targets which set in the same transmission range.

The proposed system aims to increase the lifetime of reference nodes by reducing the total amount of messages transmitted over the network. The mobile target itself requires processing its localization and transmitting it to a group leader. Therefore, it reduces the total number of messages transmitted to the sink node, and consequently reduces the total amount of energy consumed in the tracking process. Performance evaluation in terms of communication cost and power-consumption was conducted using NS2 simulator.

In this work, all of the reference nodes were considered as router nodes, which means that reference nodes have to be awake all the times, and this increase the power consumption for such WSN based tracking systems. For future work, we aim to involve the end-device nodes in the tracking process in order to minimize the power consumption, while achieving reasonable localization accuracy.

**Authors**

**Dr. Tareq Alhmiedat:** is a professor assistant in Computer Science Department at Zarqa University. He received a Ph.D. in Computer Science from Loughborough University, Loughborough, UK, 2009. MSc in Software Engineering from the University of the West of England, Bristol, UK, 2006, and a BSc degree from Applied Science University, 2004. His research interests including tracking mobile targets through Wireless Sensor Networks, Robotic Systems, and Home automation and remote control application.

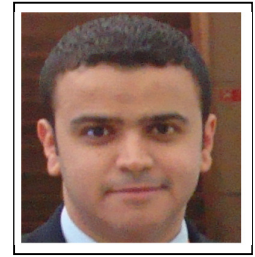

**Dr. Amer O. Abu Salem**: is a professor assistant in the department of Computer Information Systems at the Faculty of Information Technology in Zarqa University. He received his Ph.D in Computer Science from Anglia University (UK, 2010), M.Sc. in Computer Science from Al-Bayet University (Jordan, 2004), and BS.c. in Computer Science from Al-Bayet University (Jordan, 1998). Dr. Abu Salem's main research interests lie with the design and implementation of programming languages, mobile system technology, Wired and wireless communications, and program semantics.

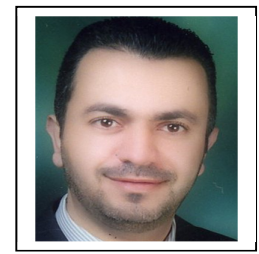

**Dr. Anas Abu Taleb**: is a professor assistant in the department of Computer Science at Isra University, Amman, Jordan. He received a Ph.D. in Computer Science from Bristol University, UK, 2010, MS.c. in Computer Science from University of the West of England, UK, 2007, and BS.c. degree in Computer Science from Prince Sumaya University for Technology, Jordan, 2004. Dr. Abu Taleb has published several journal and conference papers in sensor networks. In addition to sensor networks, Dr. Abu Taleb is interested in network fault tolerance, routing algorithms, and cloud computing.

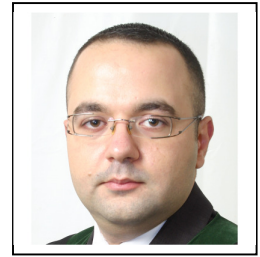